\documentclass[aps,prb,reprint,groupedaddress,floatfix,nofootinbib]{revtex4-1}
\usepackage[latin9]{inputenc}
\usepackage{amsmath}
\usepackage{amssymb}
\usepackage{amsfonts}
\usepackage{graphicx}
\begin{document}
\title{Depth profile of London length induced by nonuniform scattering rate
maximizing breakdown magnetic field in type II superconductors}
%\author{from AEK, file Optim-lam-profileTxt}
\author{A.\ E.\ Koshelev}
\affiliation{Materials Science Division, Argonne National Laboratory, Argonne,
Illinois 60439, USA}
\begin{abstract}
The breakdown magnetic field is the key parameter which determines
the performance of superconducting radio-frequency cavities. 
This is the maximum field up to which the Meissner state remains stable and  
in uniform material, it is approximately given by the thermodynamic
field. There are several recent suggestions to use nonuniform structures
to enhance the breakdown field. One of possible realizations of such
structure is depth profile of the scattering rate which, in the first
approximation, modifies the London penetration depth but does not
change the thermodynamic field. In this paper, we evaluate the optimal
profile of the London penetration depth for which the screening current density
reaches the local depairing value \emph{simultaneously at every point
within finite-size region}. Such profile is realized for close-to-linear
decrease of the London penetration depth within the length scale proportional
to its value at the surface. Achieving noticeable enhancement of the
breakdown field, however, requires strong enhancement of the London
length within large region without affecting the thermodynamic field. 
\end{abstract}
\date{\today }
\maketitle

%\section{Introduction}

%{*}{*}{*}General motivation: superconducting radio-frequency (SRF)
%cavities, breakdown (or superheating) magnetic field $B_{\mathrm{br}}$,
%reviews \cite{GurevichReviewSuST2017,PadamseeReviewSuST2017}. Recent
%experimental progress: Enhancement of performance of Nb cavities via
%nitrogen surface doping \cite{GrassellinoSuST2013,GrassellinoSuST2017}.{*}{*}{*}

Niobium-based superconducting radio-frequency (SRF) cavities became
the technology of choice in modern particle accelerators \cite{PadamseeBook2009,GurevichRevAccSciTechn2012,PadamseeReviewSuST2017,GurevichReviewSuST2017}. The cavity performance is determined by two major parameters: the quality factor $Q$ and the breakdown field $B_{\mathrm{br}}$. The latter parameter is especially important because it determines the maximum achievable accelerating gradient. 
In the simplest static picture, the breakdown field for a uniform material is determined by the metastability field of the  Meissner state which is close to the thermodynamic field $B_{c}$. At this field the current density at the surface is close to the depairing limit $j_d$.  After several decades of technological progress, the breakdown fields in modern SRF cavities are almost reached the theoretical limit for Nb, $B_c\approx 200$mT.

%In uniform material, the Meissner state remains metastable up to the
%breakdown field of the order of the thermodynamic field $B_{c}$,
%which corresponds to the current at the surface of the order of depairing current. 
There are several recent suggestions to use multilayers or nonuniform structures to improve cavity performance \cite{GurevichAIPAdv2015,KuboSuST2017,GurevichReviewSuST2017,NgampruetikornArXiv2018}. In particular, it was demonstrated that either the layer with enhanced scattering \cite{GurevichAIPAdv2015} or continuously decaying scattering rate \cite{NgampruetikornArXiv2018} may noticeably enhance the breakdown field in comparison with the uniform case. The mechanism of this positive effect is the local increase of the London penetration depth, $\lambda$, without changing of $B_{c}$. Even though this increase suppresses the local depairing current density, $j_{d}\propto B_{c}/\lambda$, it spreads the current over larger depth leading to overall increase of the maximum screening current and, consequentially, the breakdown field. In the recent work \cite{NgampruetikornArXiv2018} this effect was quantitatively demonstrated using a microscopic calculation of the static breakdown field for exponentially decaying scattering rate.
%On the other hand, 
It was discovered recently that the nitrogen injection improves the performance of niobium SRF cavities \cite{GrassellinoSuST2013,GrassellinoSuST2017}. The above ``dirty-layer'' mechanism provides a reasonable explanation for the observed positive effect on $B_{\mathrm{br}}$. In view of these developments, a natural question arises: what $\lambda$ profile would provide the maximum possible enhancement of $B_{\mathrm{br}}$? This short paper addresses this issue within the simplest theoretical framework of the linear London model.
%
%\section{Derivation}

We assume a nonuniform profile of impurity concentration leading to the coordinate-dependent mean-free path $l(x)$. Increasing scattering rate increases the local London penetration depth, in the dirty limit $\lambda(x)=\lambda_0\sqrt{\xi_0/l(x)}$ where $\xi_0$ and $\lambda_0$ are the zero-temperature clean-limit coherence and London lengths. Within the BCS scenario, the scattering does not affect neither the transition temperature nor the thermodynamic field, which is inversely proportional to the product $\lambda(x)\xi(x)$. This property is usually referred to as Anderson's theorem. To be precise, the breakdown field does not exactly coincide with the thermodynamic field and has weak nonmonotonic dependence on the scattering rate \cite{LinPhysRevB2012}. This subtle effect is beyond the scope of our consideration.

Our goal is to find the profile $\lambda(x)$ for which the screening current density reaches the local depairing value \emph{simultaneously at every points within the finite-size region}, $0<x<x_{0}$. We will evaluate such profile basing on the simplest London linear regime, within which the decay of the vector potential is described by equation
\begin{equation}
\frac{d^{2}A}{dx^{2}}-\frac{1}{\lambda^{2}(x)}A=0.\label{eq:VectPotEq}
\end{equation}
Even though the linear regime definitely breaks in the critical state,
it nevertheless is sufficient for qualitative estimates. 
In addition, the simplicity of the approach allows us to trace the physical origin of the described effect in the clearest possible way.  
The maximum local current density $j(x)=[c/4\pi\lambda^{2}(x)]A(x)$ can
not exceed the depairing current density $j_{d}(x)\propto
cB_{c}/\lambda(x)$. This means that in the critical state, in which the current density
is at the depairing level \emph{at every point}, the maximum vector
potential has to scale as $A_{\mathrm{max}}(x)\propto B_{c}\lambda(x)$.
Therefore, the profile $\lambda(x)$, for which such critical state
realizes, satisfies the following equation

\begin{equation}
\frac{d^{2}\lambda}{dx^{2}}-\frac{1}{\lambda}=0.\label{eq:OptLamEq}
\end{equation}
We assume that for $x>x_{0}$ the London penetration depth is constant,
$\lambda(x)=\lambda(x_{0})=\lambda_{0}$. 

Equation \eqref{eq:OptLamEq} has the first integral 
\[
\frac{1}{2}\left(\frac{d\lambda}{dx}\right)^{2}  =\ln\frac{\lambda}{\lambda_{\mathrm{min}}}.
\]
meaning that
\begin{equation}
\frac{d\lambda}{dx}=\pm\sqrt{\ln\frac{\lambda^{2}}{\lambda_{\mathrm{min}}^{2}}}.\label{eq:FrstInt}
\end{equation}
For fixed $\lambda$ profile, in addition to the required solution,
$A(x)\propto\lambda(x)$, the second-order equation \eqref{eq:VectPotEq}
has also another independent solution. To assure that the true dependence is indeed
$A(x)\propto\lambda(x)$ without admixture from this second solution,
it has to satisfy the continuity condition at $x=x_{0}$ for $A(x)$
and its derivative. This leads to condition 
\begin{equation}
\frac{d\lambda}{dx}(x_{0})=-1
\label{eq:BoundCond-x0}
\end{equation}
meaning that we have to select the negative sign in Eq.~\eqref{eq:FrstInt}
and the parameters $\lambda_{0}$ and $\lambda_{\mathrm{min}}$ are
connected by relation $\ln\frac{\lambda_{0}^{2}}{\lambda_{\mathrm{min}}^{2}}=1$
or $\lambda_{0}^{2}=e\lambda_{\mathrm{min}}^{2}$. 

\begin{figure}[htbp]
	\centering
	\includegraphics[width=3.4in]{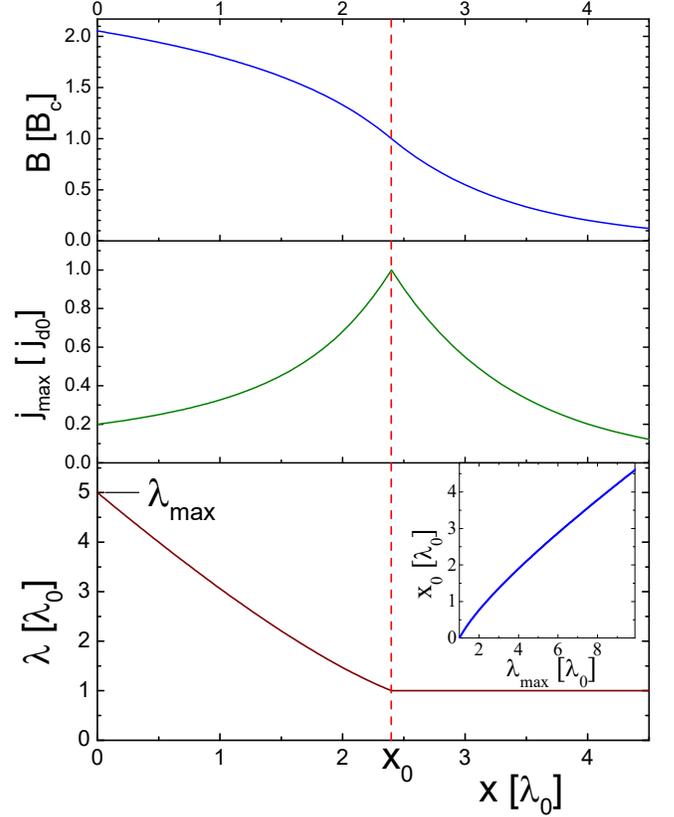}
	\caption{Lower panel: example of the optimal $\lambda$ profile for $\lambda_{\mathrm{max}}/\lambda_{0}=5$.
		Upper and middle panels show corresponding distribution of magnetic
		field and current in the critical state. The inset in the lower panel
		show dependence of the width $x_{0}$ in units of $\lambda_{0}$ on
		the ratio $\lambda_{\mathrm{max}}/\lambda_{0}$.}
	\label{Fig:LamProfCurrField}
\end{figure}

The full solution $\lambda(x)$ is implicitly determined by the integral
\begin{equation}
-\int\limits _{\lambda_{\mathrm{max}}}^{\lambda(x)}\frac{d\lambda}{\sqrt{\ln\frac{\lambda^{2}}{\lambda_{\mathrm{min}}^{2}}}}  =x,\:\mathtt{\text{for}}\,0<x<x_{0}.\label{eq:Lam-x-IntSol}
\end{equation}
with $\lambda_{\mathrm{max}}=\lambda(0)$. This implicit dependence
can also be expressed in terms of the Dawson integral $F(x)=\exp(-x^{2})\int_{0}^{x}\exp(t^{2})dt$
as 
\[
\sqrt{2}\lambda_{\mathrm{max}}F\left(\sqrt{\ln\frac{\lambda_{\mathrm{max}}}{\lambda_{\mathrm{min}}}}\right)  -\sqrt{2}\lambda(x)F\left(\sqrt{\ln\frac{\lambda(x)}{\lambda_{\mathrm{min}}}}\right)=x.
\]
The example of such dependence $\lambda(x)$ for $\lambda_{\mathrm{max}}/\lambda_{0}=5$
is shown in the lower panel of Fig.~\ref{Fig:LamProfCurrField}.
In the range $\lambda_{\mathrm{max}},\lambda(x)\gg\lambda_{\mathrm{min}}$,
using large-$x$ asymptotic of the Dawson integral, $F(x)\simeq1/2x$,
we obtain
\[
\lambda_{\mathrm{max}}\left(\ln\frac{\lambda_{\mathrm{max}}^{2}}{\lambda_{\mathrm{min}}^{2}}\right)^{-1/2}  -\lambda(x)\left(\ln\frac{\lambda^{2}(x)}{\lambda_{\mathrm{min}}^{2}}\right)^{-1/2}\approx x.
\]
corresponding to approximately linear decrease of  $\lambda(x)$. Figure \ref{Fig:LamProfCurrField} also shows the distributions of the local current $j(x)$ and magnetic field $B(x)$ for the critical state, where $j(x)\approx j_{d0}\frac{\lambda_{0}}{\lambda(x)}$ and $B(x)\approx B_{c}\sqrt{\ln\frac{e\lambda^{2}(x)}{\lambda^{2}_{0}}}$ for $x<x_{0}$ and $j(x)\approx j_{d0}\exp\left(-x/\lambda_{0}\right)$ and $B(x)\approx B_{c}\exp\left(-x/\lambda_{0}\right)$ for $x>x_{0}$.
Here $j_{d0}\propto cB_c/\lambda_{0}$ is the depairing current in the uniform region. 
Note that, in contrast to uniform state, the current reaches maximum not at the surface but inside superconductor, at $x=x_{0}$. 

As follows from Eq.~\eqref{eq:Lam-x-IntSol}, the parameters $\lambda_{0}$, $\lambda_{\mathrm{max}}$, and $x_{0}$ are connected by relation
\begin{equation}
\int\limits _{\lambda_{0}}^{\lambda_{\mathrm{max}}}\frac{d\lambda}{\sqrt{\ln\frac{e\lambda^{2}}{\lambda_{0}^{2}}}}  =x_{0}.
\label{eq:x0}
\end{equation}
or $\sqrt{2}\lambda_{\mathrm{max}}F\!\left(\!\sqrt{\ln\frac{\sqrt{e}\lambda_{\mathrm{max}}}{\lambda_{0}}}\right)
\!-\!\sqrt{2}\lambda_{0}F(1)\!=\!x_{0}$ with $\sqrt{2}F(1)\!\approx \! 0.761$. 
The plot of $x_{0}/\lambda_{0}$ vs $\lambda_{\mathrm{max}}/\lambda_{0}$ is shown in the inset of Fig.~\ref{Fig:LamProfCurrField}. In the range $2\lesssim\lambda_{\mathrm{max}}/\lambda_{0}\lesssim8$, the distance $x_{0}$ is roughly half of $\lambda_{\mathrm{max}}$. The screening efficiency can be conveniently characterized by the effective penetration depth \cite{NgampruetikornArXiv2018}
\begin{equation}
\lambda_{\mathrm{eff}}=B(0)^{-1}\int_{0}^{\infty}dxB(x),\label{eq:LamEffDef}
\end{equation}
for which we derive a closed analytical result,
\begin{equation}
\lambda_{\mathrm{eff}}=\frac{\lambda_{\mathrm{max}}}{\sqrt{1+\ln\frac{\lambda_{\mathrm{max}}^{2}}{\lambda_{0}^{2}}}}\label{eq:LamEffResult}
\end{equation}
showing that this length is mostly determined by $\lambda_{\mathrm{max}}$.

The breakdown field is proportional to the total current flowing in
the screening region,
\[
J  =\int_{0}^{x_{0}}dx\frac{cA(x)}{4\pi\lambda^{2}(x)}+\int_{x_{0}}^{\infty}dx\frac{cA(x_{0})}{4\pi\lambda_{0}^{2}}\exp\left(-\frac{x-x_{0}}{\lambda_{0}}\right),
\]
and the current enhancement with respect to the uniform case can be
estimated as 
\begin{figure}[htbp]
	\centering
	\includegraphics[width=3.4in]{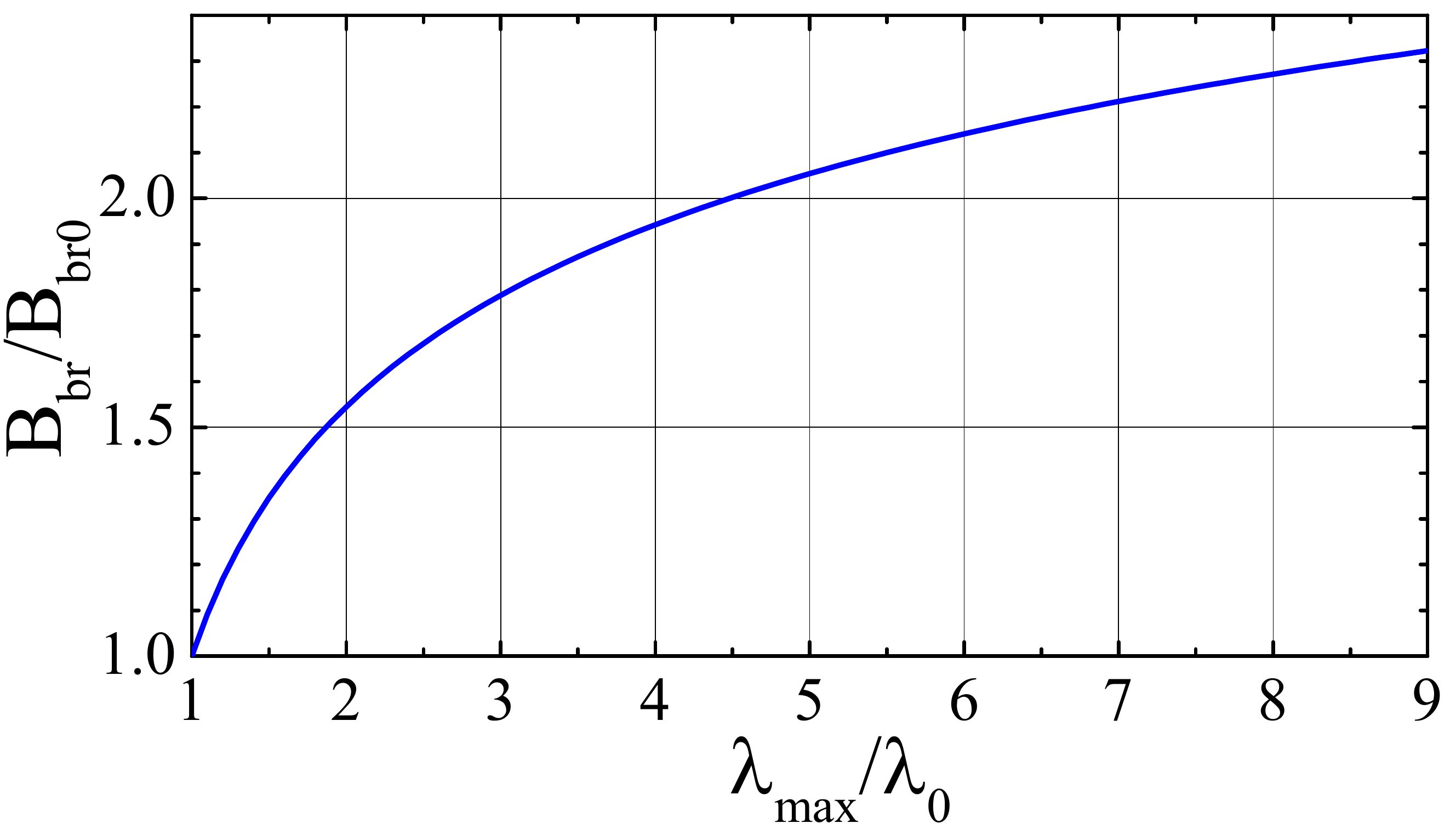} \caption{The dependence of the breakdown field enhancement on the ratio $\lambda_{\mathrm{max}}/\lambda_{0}$. }
	\label{Fig:BbrEnh}
\end{figure}
\begin{equation}
\frac{J}{J_{\mathrm{u}}}  \!=\!\int_{0}^{x_{0}}\!\frac{dx}{\lambda(x)}+1=\frac{d\lambda}{dx}(x_{0})-\frac{d\lambda}{dx}(0)+1\!=\!-\frac{d\lambda}{dx}(0),\label{eq:TotalCurrEnh}
\end{equation}
where the derivative of $\lambda$ is connected with its value at the
surface, Eq.~\eqref{eq:FrstInt}. Therefore, the breakdown field
enhancement with respect to the uniform case can be represented as
\begin{equation}
\frac{B_{\mathrm{br}}}{B_{\mathrm{br0}}}=\sqrt{1+\ln\frac{\lambda_{\mathrm{max}}^{2}}{\lambda_{0}^{2}}}.
\label{eq:TotalCurrEnhRes}
\end{equation}
This dependence is plotted in Fig.~\ref{Fig:BbrEnh}. Formally, within this simple model, the increase of $B_{\mathrm{br}}$ can be arbitrarily large. This increase, however, is a very slow function of the ratio $\lambda_{\mathrm{max}}/\lambda_{0}$ and significant enhancement requires controlled strong suppression of $\lambda$ within a wide region. For example, doubling of $B_{\mathrm{br}}$ requires increase of $\lambda$ at the surface by the factor $\sim 4.5$. %

The result in Eq.\ \eqref{eq:TotalCurrEnhRes} can be compared with the breakdown field enhancement for the \emph{uniform} dirty layer \cite{GurevichAIPAdv2015,GurevichReviewSuST2017}. In this case for the optimal thickness of this layer, the screening currents reaches the depairing values simultaneously at the 
boundary of the dirty layer with free space and at the boundary between the clean and dirty materials.  The breakdown field enhancement can be evaluated as $B_{\mathrm{br}}/B_{\mathrm{br0}}=\sqrt{2-\lambda_0^2/\lambda^2}$, where $\lambda_0$ and $\lambda$ are the London lengths of the clean material and dirty layer respectively 
%\footnote{This result follows from Eq.\ (7) of Ref.\ \cite{GurevichAIPAdv2015} for $H_s=H_s0$, which is $B_{\mathrm{br0}$ in our notations.}. 
We see that in this case the maximum enhancement for $\lambda\gg \lambda_0$ is $\sqrt{2}$. Going from uniform to distributed dirty layers allows to reach higher enhancements.  

To what extend the London length can be practically increased without affecting the thermodynamic field is not well known and deserves thorough investigation. An obvious theoretical limit is set by the condition $k_Fl> 1$ giving $\lambda_{\mathrm{max}}<\lambda_{0}\sqrt{k_F\xi_0}\approx \lambda_{0}\sqrt{\epsilon_F/\Delta}$, where $k_F$ is the Fermi momentum, $\epsilon_F$ is the Fermi energy, and $\Delta$ is the superconducting gap.

In conclusion, we evaluated profile of the London length for which the critical state is achieved simultaneously within a finite region. For such profile, the significant enhancement of the breakdown field may be achieved but requires large increase of the London length within wide region. The evaluation is based on the simplest linear London theory and has to be further verified by more accurate theoretical models.

This work has been initiated and motivated by discussions with Vudtiwat Ngampruetikorn, James Sauls, Ulrich
Welp, Wai K. Kwok, Leonardo Civale, Andreas Glatz, Anna Grassellino, and Alexander
Romanenko. This work was supported by the U.S. Department of Energy,
Office of Science, Basic Energy Sciences, Materials Sciences and Engineering
Division.

\bibliography{SRF}

\end{document}